\documentclass[a4paper]{jpconf}
\usepackage{graphicx}
\usepackage[sort&compress,square,comma,numbers]{natbib}

\begin{document}
\title{Superconductivity in the Kondo lattice model}
\author{O Bodensiek$^1$, T Pruschke$^1$, R \v{Z}itko$^{1,2}$}
\address{$^1$ University of Goettingen, 37077 Goettingen, Germany\\
$^2$ J.\ Stefan Institute, Jamova 39, SI-1000 Ljubljana, Slovenia}
\ead{pruschke@theorie.physik.uni-goettingen.de}
\begin{abstract}
We study the Kondo lattice model with additional attractive interaction
between the conduction electrons within the dynamical mean-field theory
using the numerical renormalization group to solve the effective
quantum impurity problem. In addition to normal-state and magnetic  
phases we
also allow for the occurrence of a superconducting phase.
In the normal phase we observe a very sensitive
dependence of the low-energy scale on the conduction-electron  
interaction.
We discuss the dependence of the superconducting transition on the  
interplay between
attractive interaction and Kondo exchange.
\end{abstract}
\section{Introduction}
Heavy Fermion (HF) compounds with elements from the lanthanide or  
actinide
series share some rather general features in the Fermi liquid phase,  
namely
a strong enhancement of the effective carrier mass and a similarly  
enhanced
Pauli susceptibility with a Wilson ratio typically of the order, but  
larger
than one \cite{stewart:84,GreweSteglich:1991}.
In addition to this well-understood Fermi liquid phase
\cite{GreweSteglich:1991,hewson:book,niki:97PAM,pruschke:00,Grenzebach:2006},
HF systems also exhibit a variety of phase transitions, among them  
magnetic and
superconducting phases.
This aroused the strong interest of both experimentalists and  
theorists, as
$f$-electrons conventionally tend to suppress superconductivity. The
discovery of quantum critical phenomena
\cite{Stewart:2001,Stewart:2006Add,Loehneysen:2007} eventually showed  
the
intimate link between the latter two.
In spite of the rather large collection of experimental results, an  
accepted
theoretical description of superconductivity has not yet been  
established.
Moreover, the role of phonons on the low-energy properties of HF  
compounds
and in particular their relevance for a microscopic theory of
superconductivity in HF systems has not yet been addressed in detail
\cite{Grewe:1984SC,Grewe:1984SC:2,Barzykin:2005}.

We present the first results of a study of the Kondo lattice model
with an effective attractive interaction among the conduction  
electrons. The
latter can be thought to be obtained from an optical phonon mode  
treated in
the antiadiabatic limit (a more realistic description employing a true
optical phonon in the calculation is the subject of ongoing  
investigations).
Our model is thus
\begin{eqnarray}
   \label{eq:1}
   H &=& \sum\limits_{i,j,\sigma} t_{ij,\sigma}c^{\dagger}_{i\sigma}
c^{\phantom{\dagger}}_{j\sigma}+U\sum\limits_{i}
c^{\dagger}_{i\uparrow}c^{\phantom{\dagger}}_{i\uparrow}c^{\dagger}_{i 
\downarrow}c^{\phantom{\dagger}}_{i\downarrow}
-\frac{J}{2}\sum\limits_{i,\alpha\beta}
\vec{S}_i\cdot\vec{\sigma}_{\alpha\beta}\,c^{\dagger}_{i\alpha}
c^{\phantom{\dagger}}_{i\beta}
\end{eqnarray}
where $c^{(\dagger)}_{i\sigma}$ annihilates (creates) a conduction  
electron at lattice site
$\vec{R}_i$ with spin $\sigma$, $U$ is the effective interaction  
between conduction electrons,
and $J$ the Kondo exchange. Note that in our notation antiferromagnetic
coupling means $J<0$. Finally, $\vec{\sigma}$ denotes the vector of  
Pauli
spin matrices. We solve the model with dynamical mean-field theory  
(DMFT)
\cite{georges:96} and Wilson's numerical renormalization group (NRG)
\cite{bullareview}. Calculations were performed for a Bethe lattice with
infinite coordination number.
In order to study superconductivity we allow for a corresponding  
symmetry
broken phase \cite{Bauer:2009}. Note that we cannot include  
unconventional
order parameters here, as DMFT only allows for $s$-wave phases
\cite{georges:96}.

\section{Results}
\subsection{Paramagnetic phase}

\begin{figure}[htb]
   \centering
   \includegraphics[width=0.65\textwidth,clip]{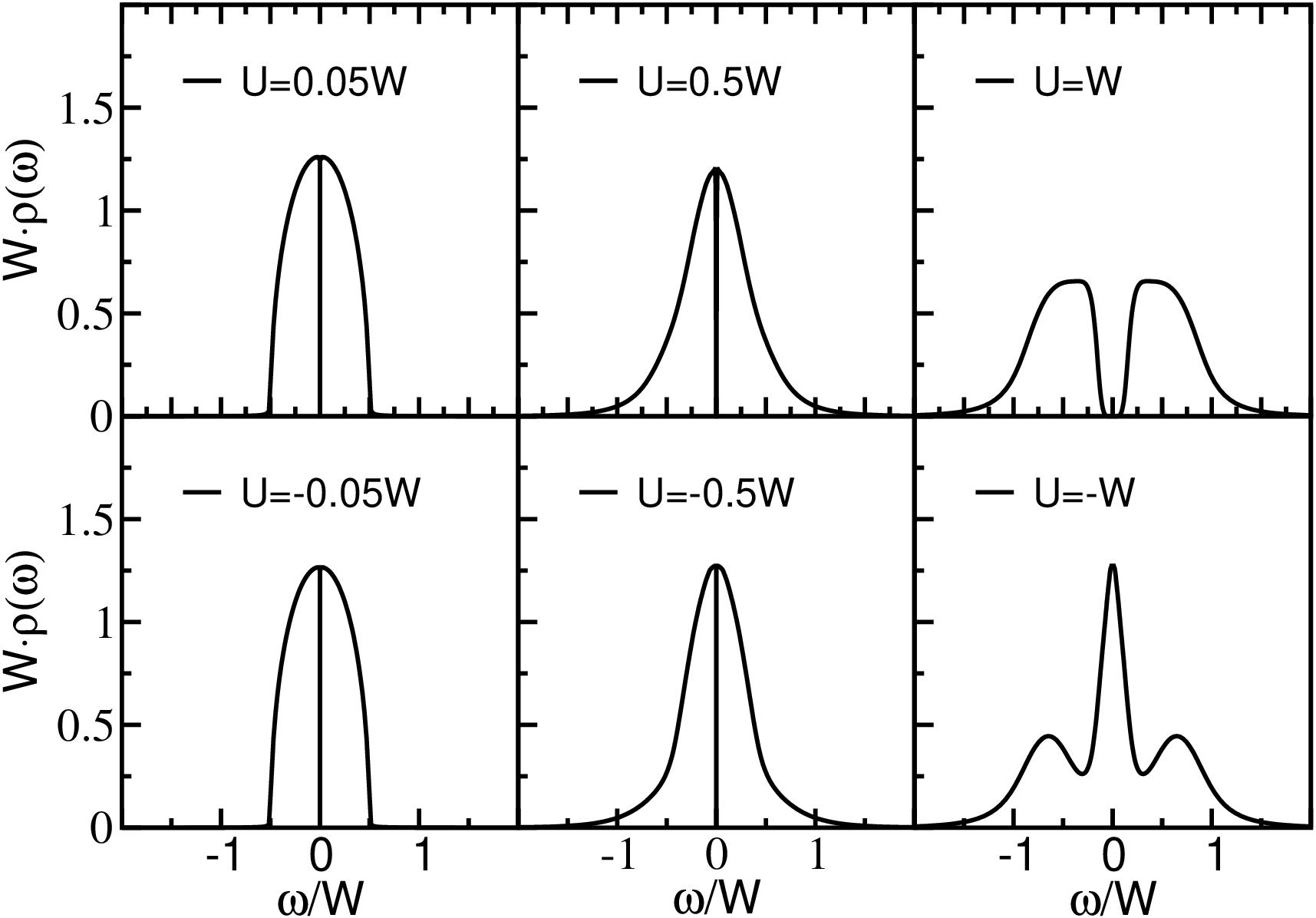}
\hfill
   \includegraphics[width=0.33\textwidth,clip]{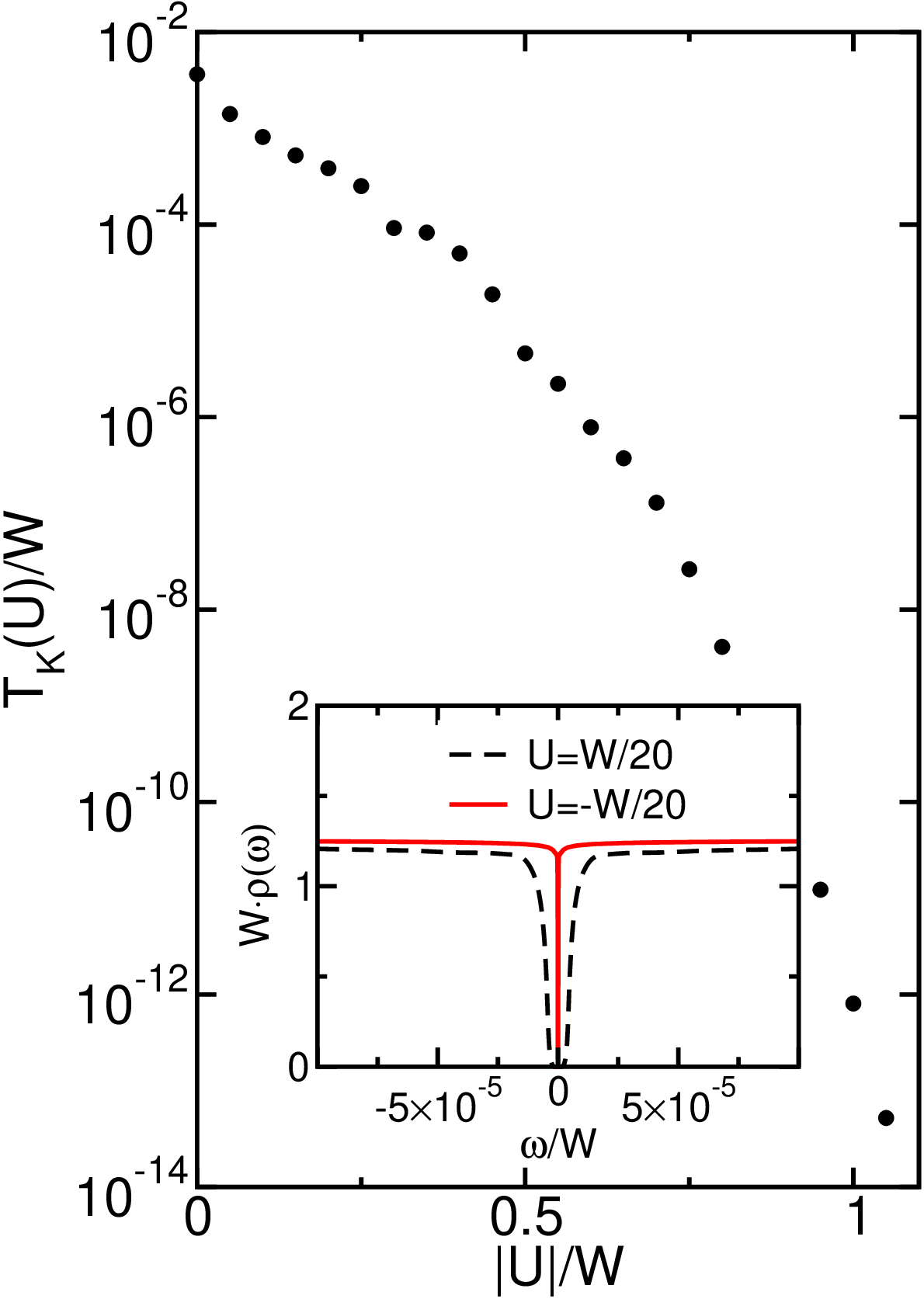}

   \caption{Left: Density of states for the Kondo model (\ref{eq:1})  
at half
filling with $J=-W/16$  at $T=0$.
Right: Low-energy scale obtained from the width of the gap in the  
attractive case
$U<0$ for $J=-W/16$. The inset shows the different gaps for small  
repulsive
and attractive
$U$.
 }
   \label{fig:1}
\end{figure}

It is necessary to stress that the model (\ref{eq:1}) does not show the
usual symmetry $U\leftrightarrow-U$ under simultaneous exchange of  
spin and
charge, i.e.\ its physics cannot be inferred from the corresponding  
magnetic
properties of the model with repulsive interaction.
In Fig.\ \ref{fig:1} we compare the two cases $U>0$ and $U<0$.
Superficially, for weak interaction $|U|<W$,
both seem to be rather similar. However, as is evident from the inset  
to the right part and the behavior for larger interaction
the insulator is much stronger for repulsive $U$. It here
originates from the formation of a local spin singlet rather than  
being of
Mott-Hubbard type \cite{Peters:2007}. For attractive $U<0$, on the other
hand, we find that Kondo screening is strongly suppressed and the system
eventually recovers Mott-Hubbard physics in the charge sector for large
$|U|$. This difference can be easily understood. For attractive $U$ the
conduction system namely experiences strong correlations in the charge
sector, while the Kondo exchange tries to develop such feature in the  
spin
sector. Obviously, when $|U|>T_{\rm K}^{(0)}$, where $T_{\rm K}^{(0)} 
$ is
the Kondo scale for $U=0$, spin fluctuations will efficiently be  
suppressed,
leading to the observed behaviour. The suppression of the Kondo scale is
actually stronger than exponential, as shown in the right part of  
Fig.~\ref{fig:1}.

From these results we draw two conclusions: 1) phonons are clearly  
extremely
important even to the paramagnetic phase of the Kondo model and thus to
properly account for the low-energy scale of HF systems; 2) as the
low-energy scale is efficiently reduced by an attractive interaction  
among
the conduction electrons, we expect that $s$-wave superconductivity will
actually prevail in a large part of the phase diagram, in particular for
small Kondo coupling $J$.

\subsection{Superconducting phase}
In order to allow the system to show superconductivity, we have to  
reformulate the
DMFT equations in Nambu space and extend the NRG accordingly. The  
latter has been
accomplished some time ago already (for a review see \cite 
{bullareview} and references
therein; the actual way to combine DMFT and NRG has been extensively  
discussed by Bauer \cite{Bauer:2009}).

\begin{figure}[htb]
   \centering
   \includegraphics[width=0.65\textwidth,clip]{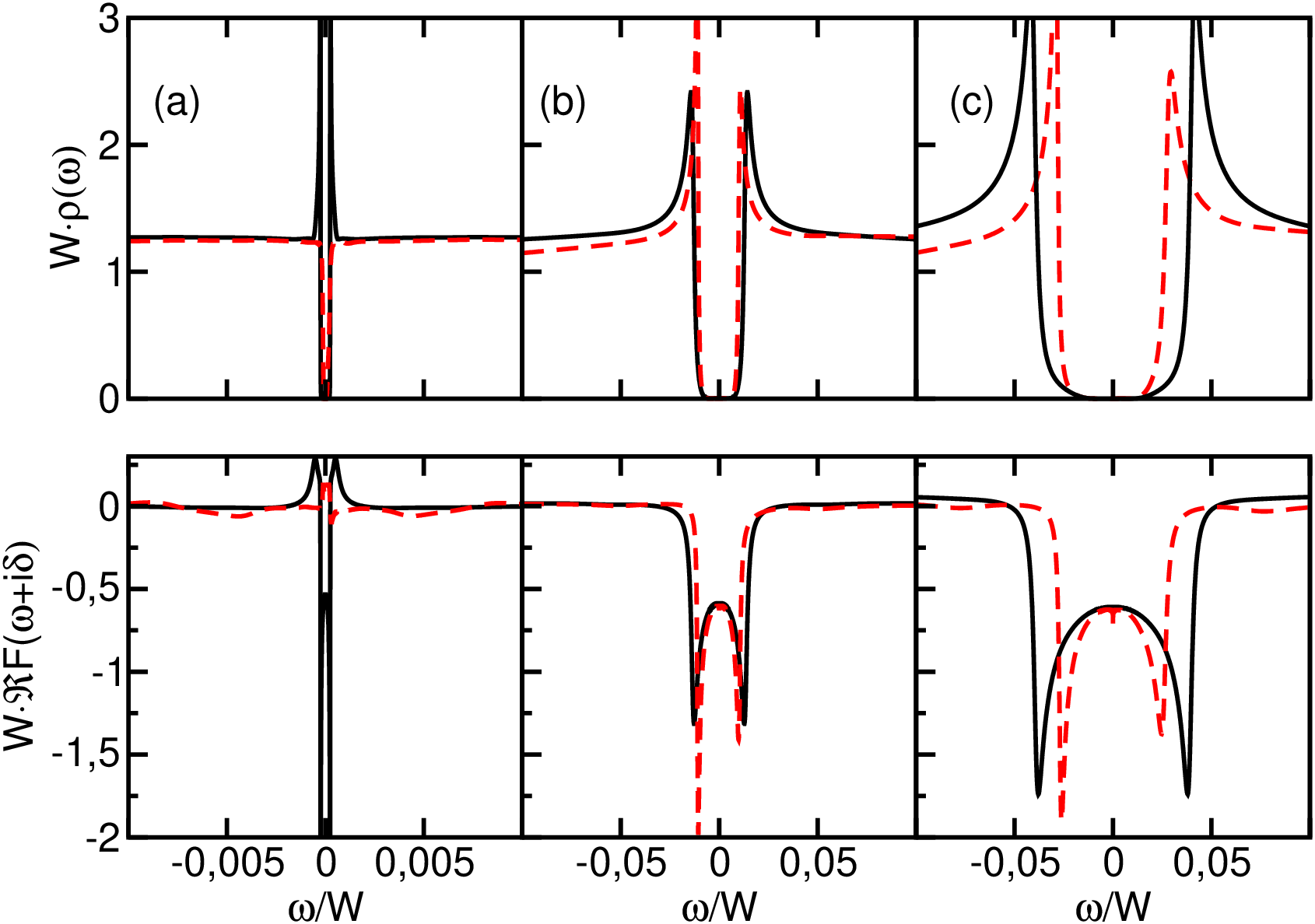}
   \caption{DOS (upper panels) and real part of the anomalous Green's
function (lower panels) for small Kondo exchange $J=-W/25$ for (a) 
$U=-W/80$, (b) $U=-W/8$ and (c) $U=-W/4$.
  }
   \label{fig:3}
\end{figure}
For a very small Kondo exchange interaction $J=-W/25$, the ground  
state of
the model is dominated by superconductivity. This becomes apparent from
Fig.\ \ref{fig:3}, where the DOS (upper panels) and the real part of the
anomalous Green's function (lower panels) is shown at half filling (full
curves) as well as at finite filling $\langle n\rangle\approx0.75$  
(dashed
curves). As the low-energy scale of the model with $U=0$ is always  
largest
at half filling \cite{pruschke:00}, we can expect that this result
remains stable for all fillings $\langle n\rangle\le1$. Note that in  
none of
the cases one does observe a significant dependence of the gap on the
fillings, i.e.\ local correlations due to Kondo screening are frozen out
here since $T_{\rm K}\ll |U|$. Furthermore, $|U|/W<1$ and  
consequently one
expects and observes a BCS like gap structure, only weakly smeared  
out by
self-energy broadening.

\begin{figure}[htb]
   \centering
   \includegraphics[width=0.65\textwidth,clip]{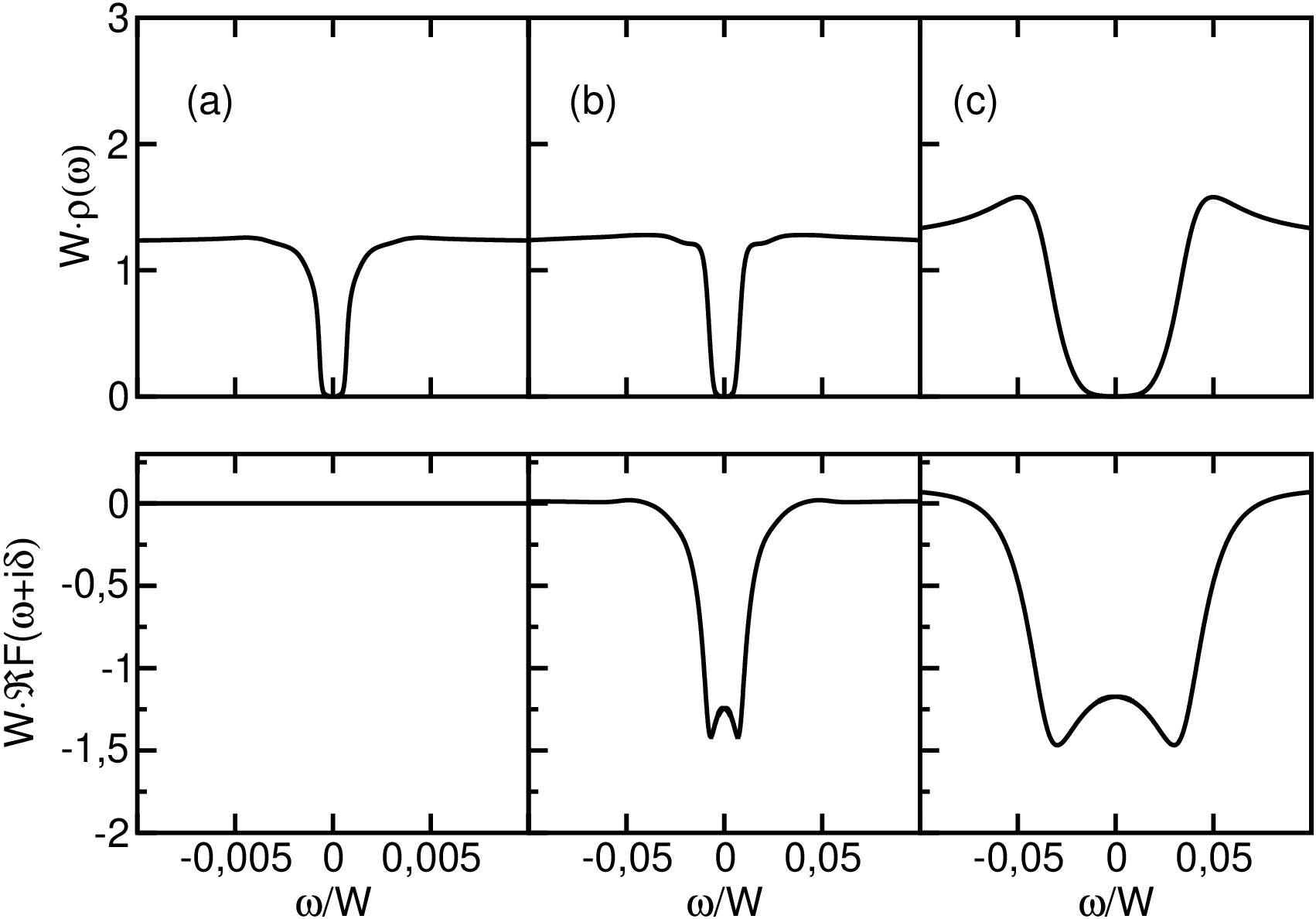}
   \caption{DOS (upper panels) and real part of the anomalous Green's  
function (lower panels)
   for larger Kondo exchange $J=-W/10$ and (a) $U=-W/100$, (b) $U=-W/10$ and
(c) $U=-W/5$.}

  \label{fig:4}
\end{figure}
Increasing $|J|$ has two effects. First, there appears a finite,  
critical $U_c$ below which
no superconducting solution exists. This can be seen in Fig.\ \ref 
{fig:4}a, where DOS
and real part of the anomalous Green's function are shown for $J=-W/10 
$ and a small
$U=-W/100$. Note that at half filling we find a Kondo insulator,  
which has a gap in
the DOS. From that perspective the result is actually  
indistinguishable from the superconducting phase. The anomalous part,  
however, vanishes here, i.e.\ we have indeed a normal state and thus  
an insulator. For larger interactions, the superconducting phase  
reappears. Compared to the case with small $J$, we observe here  
visible reduction of the gap and also a broadening of the  
singularities at the gap edges.
We attribute this behavior to the correlations induced by the Kondo  
exchange.

\section{Conclusion} We have studied the Kondo lattice model with an
attractive interaction between the conduction electrons,
which may arise in the presence of phonons, notably optical modes like
breathing modes. We found a tremendous effect of an attractive  
interaction
on the low-energy scale, drastically reducing it already for  
comparatively
modest $|U|$. This behavior can be understood in terms of a competition
between a spin Kondo effect and the charge fluctuations introduced by  
the
attractive $U$. As the latter also favour superconductivity we can  
expect,
and indeed do find, that for experimentally relevant values of the Kondo
coupling and interaction parameters the model shows an $s$-wave type
superconducting ground state. For real systems this underlines the
importance of elastic degrees of freedom for a proper description of the
physics of HF materials.

\end{document}